\documentclass{article}
\usepackage{iclr2021_conference,times}

\usepackage[utf8]{inputenc}
\usepackage[english]{babel}
\usepackage{url}
\usepackage{mathtools}
\usepackage{amsfonts}
\usepackage{hyperref}
\hypersetup{
colorlinks=true,
linkcolor=blue,
urlcolor=cyan,
citecolor=magenta
}

\usepackage{epigraph}
\usepackage{vhistory}
\usepackage{subcaption}
\usepackage{grffile}
\usepackage{algorithm}
\usepackage{algpseudocode}
\usepackage{algorithmicx}
\usepackage{multirow}
\usepackage[normalem]{ulem}
\usepackage{mathtools}

\def\eqref#1{equation~\ref{#1}}
\def\1{\bm{1}}

\DeclareMathAlphabet{\mathsfit}{\encodingdefault}{\sfdefault}{m}{sl}
\SetMathAlphabet{\mathsfit}{bold}{\encodingdefault}{\sfdefault}{bx}{n}

\newcommand{\E}{\mathbb{E}}

\DeclareMathOperator*{\argmax}{arg\,max}

\newcommand{\regret}{g}

\title{Learning to Rank For Push Notifications Using Pairwise Expected Regret}

\author{Yuguang Yue\thanks{These authors contributed equally.}, Yuanpu Xie, Huasen Wu, Haofeng Jia, Shaodan Zhai, Wenzhe Shi, Jonathan J Hunt$^*$
\\
\texttt{\{yuguangy, yxie, huasenw, hjia, szhai, wshi, jjh\}@twitter.com}
}

\iclrfinalcopy

\begin{document}

\maketitle

\begin{abstract}
Listwise ranking losses have been widely studied in recommender systems. However, new paradigms of content consumption present new challenges for ranking methods. In this work we contribute an analysis of learning to rank for personalized mobile push notifications and discuss the unique challenges this presents compared to traditional ranking problems. To address these challenges, we introduce a novel ranking loss based on weighting the pairwise loss between candidates by the expected regret incurred for misordering the pair. We demonstrate that the proposed method can outperform prior methods both in a simulated environment and in a production experiment on a major social network.
\end{abstract}

\section{Introduction}

The majority of internet users now access the internet via a mobile device \citep{handley2019mobile}. As a result, mobile devices have become increasingly important for content consumption.
Push notifications allow a mobile user to receive timely messages and relevant information from an app they have installed, even while the app is in the background.

Because push notifications interrupt the user flow, users have a low tolerance for irrelevant content and expect to receive notifications only for content that is timely and important to them. This makes ranking and decision making for push notifications crucial to the user experience and a demanding problem.

Learning to rank is an important topic in information retrieval for designing machine learning ranking systems to surface relevant content to users from a large corpus \citep{liu2011learning}. These ranking models are typically learned using logged user feedback on previous content.
Listwise ranking is an approach to modeling the ranking problem with a loss which more closely approximates the utility of a ranking over documents to a user. For example, in a traditional ``10 blue links'' search paradigm \citep{chen2012beyond}, the user is only shown the top 10 results. Therefore, the ordering of documents after the 10th will have no impact on the utility of the ranking, a property a listwise ranking loss can exploit.
Many metrics and listwise ranking losses implicitly focus on this search paradigm.

In this work, we present the ranking problem for push notifications and its unique challenges on a the Twitter platform. While some specifics may be unique to the platform, the problem constraints will be similar in other content-based push settings.

We observe that push notification differ from many classical ranking problems in several key ways. 1. Unlike other ranking problems such as search, the user will observe only one notification from each set of candidates. 2. User behavior is highly personalized since there is no explicit context from the user to indicate their information need. 3. User responses are highly non-stationary, a notification that is timely and relevant now, may be irrelevant if sent later. 4. The candidates available to be sent to the user changes rapidly and is highly personalized, so the approach must generalize to rank documents never seen before.
These characteristics combine to make counterfactual estimates challenging. See section \ref{sec:char_push} for more details.

We derive a ranking loss which approximately minimizes the expected regret of the ranking in the push notification setting by weighting pairwise losses. We test this approach in both a simulation based on the production system and a production experiment on real users. We show that the proposed loss can outperform prior approaches and results in significant gains over baselines in a production AB test.

\section{Related work}\label{sec:related_work}

The ranking problem is to sort a candidate set of documents by relevance to the user. There has been a significant body of work studying ranking losses and metrics known as learning to rank \citep{oosterhuis2020unbiased}. A number of metrics have proposed to estimate performance of a ranking. For example, the widely used Normalized Discounted Cumulative Gain \citep{jarvelin2002cumulated}, Expected Reciprocal Rank \citep{chapelle2009expected}, and
Average Relevance Position \citep{Zhu2004RecallPA}. These ranking metrics are non-differentiable and therefore not practical to optimize directly. Their surrogate functions can be optimized via gradient based algorithms \citep{cao2007learning,xia2008listwise}.

\textit{Pairwise losses} are a  class of ranking losses that compare positive and negative pairs to minimize the number of inversions \citep{chen2009ranking}. In practice, \citet{pasumarthi2019tf} empirically demonstrates that \textit{listwise} losses typically outperform \textit{pairwise} losses on a range of tasks, because of \textit{pairwise} losses do not take ranking information into account.

Many listwise losses can be decomposed into weighted pairwise losses. Well known ranking losses include RankNet \citep{burges2005learning}, LambdaRank \citep{burges2006learning} which improves on RankNet by adding weights to the gradient based on the cost of inversions, a tree-based LambdaMART \citep{wu2010adapting}.

To take advantage of the ranking information, \citet{usunier2009ranking} proposed the ordered weighted pairwise classification (OWPC) algorithm. However, as with many \textit{listwise} losses (e.g.\   \textit{ListNet}\citep{xia2008listwise} and $SVM_{map}$ \citep{yue2007support}), OWPC requires knowledge of the rank of positive examples to calculate the loss, which is prohibitively expensive when the candidate set has millions of items. To make this idea more scalable \citet{weston2013learning} proposed the $K$-$Order\ Statistics\ Loss$ (K-OS) loss which is an extended loss function of the weighted approximately ranked pairwise loss \citep{weston2011wsabie}; K-OS loss takes care of both the estimation of rank and reweighting the pairwise loss. Weighted Margin-Rank Batch loss proposes a similar method for the batch training regime \citep{liu2017wmrb}.

In addition to the learning to rank approaches to push notifications, two recent papers \citep{yancey2020sleeping,xu2020contextual} proposed contextual bandit based algorithms for push notification and personalized recommendation problems, where the arm of the bandit is interpreted as the item to be sent. However, both algorithms require repeated arms, which are not available in our setting where content creation happens rapidly.

\section{Push notification ranking problem}
\label{section_push_notification_ranking_problem}

Changes in technology have resulted in new modes of users consuming content. In particular, on mobile devices users can choose to receive ``push notifications'' to be alerted to relevant new content in a timely manner. We describe the basic setup of the recommender based notification system at Twitter.

Algorithm \ref{algo:push} shows pseudocode of the notification system. $\mathcal{U}$ denotes the set of users and $\mathcal{X}$ the space of features of documents. Periodically a pass is made through the set of users, candidate documents for each user are obtained and a scoring function, parametrized by $\theta$, $f_{\theta}(u, x): \mathcal{U} \times \mathcal{X}  \rightarrow \mathbb{R}$ is used to score the candidates.
%Provided the score of the highest-ranked document is above some user-dependent threshold, it is sent to the user.
The highest scoring document is sent to the user and the user response (which is binary, either the user opens the notification $y=1$ or ignores it $y=0$) is logged and used to train subsequent ranking models.

\begin{algorithm}[h]
  \begin{algorithmic}
    \State Accept scoring function $f_\theta(u, x)$
    \For{$u \in \mathcal{U}$}
      \State Obtain candidate documents $\{x_1, \cdots x_n\}$ for user $u$ available this time.
      \State Find the highest scoring document $x = \argmax_{{x} \in \{x_1, \cdots x_n \}} {f_{\theta}(u, x)}$
      \State Send the corresponding document to the user and receive a response $y \in \{0, 1\}$
      \State Log $u, x, y$
    \EndFor
  \end{algorithmic}
  \caption{Pseudo-code of a single pass of the ranking based push notification system. At each timestep a set of candidates for a user are obtained (note this list will be distinct for each user and each pass through the loop due to new content being created), ranked and the highest scoring candidate sent to the user.
  }
  \label{algo:push}
\end{algorithm}
The objective of the ranking system is to surface documents relevant to the user, as measured by the user choosing to open the notifications rather than dismiss them.

\subsection{Characteristics of push notifications}\label{sec:char_push}

Push notifications have characteristics that make ranking challenging and distinct from many existing ranking problems. Firstly, of all the candidates ranked, only a single candidate can be sent to the user and receive feedback due to the display limitations of push notifications. This is important when considering what is the appropriate ranking loss below, and distinct from many other ranking problems.

Secondly, document relevance is highly personalized. Unlike some other content consumption routes, users receive push notifications without actively interacting with an app, so there is limited context (e.g.\ in a search engine if several users search for the same keywords, their responses could be grouped together). This high degree of personalization means that it is not valid to average user responses to the same document.

Thirdly, user responses are non-stationary. That is a document may be relevant to the user now, but not relevant a short time in the future.
For example, users make use of Twitter to obtain information on breaking news \citep{petrovic2013can, osborne2014facebook}. A document sent to a user now may be opened, the same document sent later may be irrelevant and dismissed later.

Finally, new documents are created at high volume. Therefore, ranking approaches must generalize to documents that have not been seen before. Each time a set of candidates is ranked, both this set and many individual candidates in the set may have never been seen by the system before.

These properties combine to make it challenging to estimate counterfactual outcomes. In contrast, in a search engine where the user views multiple results, after correcting for position bias \citep{craswell2008experimental}, it is then possible with some assumptions to estimate the performance of a ranking with a different ordering of the same documents \citep{agarwal2019general}. The personalization and non-stationarity problems combine to mean that we can't average multiple user responses together to get better relevance estimates than binary labels in contrast to typical contextual search problems \citep{joachims2002optimizing}.

Like most ranking problems, we could also frame this as a contextual bandit problem as (see section \ref{sec:related_work}). However, because of the rapid creation of new documents, standard bandit approaches would not be relevant since each arm of the bandit would only be seen once and we need to generalize to unseen arms. A contextual bandit in our problem setup would reduce to a pointwise loss. For this reason, in this work we have focused treating the problem as a ranking problem.

\subsection{Push notification ranking utility} \label{sec:pushutility}

In this work we deal only with deterministic policies. Because in push notifications, only the top ranked document is sent to the user and all other documents have no effect on the outcome, we define a policy $\pi: \mathcal{U} \times \{\mathcal{X}_{1}, \cdots,\mathcal{X}_{n}\}$ as a mapping from a user $u$ and unordered candidate set of documents $\{x_1, \cdots, x_n\}$ to one document in that set $x$. Note that the ordering over documents in the set is arbitrary.
The chosen document $x = \pi(u, \{x_1, \cdots, x_n\}) $ results in a reward $r(u,x) =y$ based on the response of the user $y \in \{0, 1\}$, who may either open the document $y=1$ or dismiss it $y=0$.

The reward is stochastic, as there are many unobserved factors which affect the outcome (e.g.\ a relevant notification sent when the user is busy may be dismissed). In simulation we model each document as having a latent likelihood of being opened if it is sent to the user $\tilde{y} = p(y=1|u, x)$ 
which defines a Bernoulli distribution. In the simulation, where we have access to the latent $p(y=1|u, x)$ for each document, we also define a deterministic reward function $r_{sim}(u, x) = p(y=1|u, x)$. Note that $r_{sim} = \E_{p(y|u, x)} r(u,x)$. This latent variable is never used in training (that would be unrealistic), but allows us to evaluate the ranking algorithms with lower variance in simulation.

The regret (which we can only compute exactly in simulation when we have access to the latent properties of the document) incurred by a ranking policy $\pi$ for a particular user $u$ and candidate set of documents ${x_1, \cdots, x_n}$ is given as:
\begin{align}
\regret(u, \{x_1, \cdots, x_{n}\}, \pi) =  \max_{x_j \in \{x_{1}, \cdots, x_{n}\}} r_{sim}(u, x_j) - r_{sim}(u, \pi(u, \{x_1,\cdots,x_n\})).
\end{align}

The way we actually parametrized our ranking policy for all methods here is by a  scoring function $f_{\theta}: \mathcal{U} \times \mathcal{X} \rightarrow \mathbb{R}$ that maps from user features and document features to a real-valued score, $f_{\theta}$ is parameterized by $\theta$. %\yy{Can we explicitly state the relation between scoring function and policy e.g., $\pi(u,d) =d_{top} \triangleq  d_{\argmax_i f_{\theta}(u,x_i)}$ ?}
The highest scoring document is selected $\pi(u, \{x_1, \cdots, x_n\}) = \argmax_{x \in {\{x_1, \cdots, x_n\}}} f_{\theta}(u, x)$.

\subsection{User types}

Users at Twitter are categorized into 7 different user types based on their usage of the platform. The behavior of users to push notifications varies by user type (e.g.\ a user who uses Twitter every day is more likely to open a notification than an occasional user). We made use of the user type in our ranking method described below.

\section{Method}
\label{sec:method}

Here we describe both the prior methods for ranking that we use as baselines and introduce a novel loss designed for ranking push notifications.

We write all the losses over a single candidate set. We split the set of candidates into candidates with positive labels (opened) $\mathcal{X}_{pos}$ and candidates with negative labels (dismissed notifications) $\mathcal{X}_{neg}$.
We assume the final output of the model $f_\theta(u, x)$ has no non-linearity and we write any final non-linearity in the loss itself. 

\subsection{Existing ranking losses}\label{existing_loss}

\subsubsection{Pointwise}

A straightforward approach, widely used in practice \citep[e.g.\ ][]{liu2017related}, is to treat the ranking problem as a classification problem on the outcome $y$ and use the cross-entropy loss to train the score function with empirical risk minimization to predict the likelihood that a document will be opened. The loss is: 
\begin{align}
    \ell_{ce}(u, \theta) = \sum_{x_{pos} \in \mathcal{X}_{pos}} - log(\sigma(f_{\theta}(u, x_{pos})) - \sum_{x_{neg} \in \mathcal{X}_{neg}} \log(1 - \sigma(f_\theta(u, x_{neg}))
\end{align}
where a sigmoid function $\sigma$ is used to bound the prediction $f_{\theta}(u, x)$ to $(0, 1)$.

The well-known weakness of a pointwise loss is that it is a poor proxy of the true objective of the ranking function \citep{liu2011learning} because of the mismatch between the objective of predicting the outcome for each document, and the ranking problem in which only the score relative to other documents in the candidate set are important.
This is particularly acute in the push notification setting where only a single item will be sent to the user and therefore only the item which is ranked highest matters for the outcome.

\subsubsection{Pairwise}

Pairwise losses focus on the relative ordering of two documents (with different outcomes). The pairwise hinge loss is
\begin{align} \label{eq:pairwise1}
    \ell_{pair}(u, \theta) =\sum_{x_{pos} \in \mathcal{X}_{pos}} \sum_{x_{neg} \in \mathcal{X}_{neg}} \max(0, 1 - (f_{\theta}(u, x_{pos}) - f_{\theta}(u, x_{neg})))
\end{align}
The hinge loss ``pushes'' the score function to score documents with a positive label higher than negative documents. There are other pairwise losses that could be used, such as a pairwise logistic loss, but in initial experiments we found a hinge loss performed best.

\subsubsection{Listwise ranking losses}

Listwise ranking losses try to approximate the utility of the ordering of the set more directly. The closest prior example to the loss we introduce below is the  $K$-$Order\ Statistics\ Loss$ (K-OS) AUC loss \citep{weston2013learning} which incorporates ranking information into a pairwise loss.

K-OS first computes an ordering over the positive examples based on the current scores of the model being optimized $(x_{pos}^1, \cdots, x_{pos}^{|\mathcal{X}_{pos}|})$.

\begin{equation} \label{eq:kos}
    \ell_{K-OS-AUC}(u, \theta) = \frac{1}{Z}\sum_{i=1}^{|\mathcal{X}_{pos}|}W(\frac{i}{|\mathcal{X}_{pos}|}) \sum_{x_{neg} } \max(0, 1-(f_{\theta}(u, x^{i}_{pos})-f_{\theta}(u,x_{neg})))
\end{equation}
where $Z = \sum_i W(\frac{i}{|\mathcal{X}_{pos}|})$. $W$ weights pairs by their rank. If $W(j) = C$ for all $j$ and $C$ is a positive constant, K-OS loss just reduces to the pairwise loss; if $W(i)>W(j)$ when $i<j$, it will penalize errors on high ranked positive items more.
In our setup, $W(\frac{i}{|\mathcal{X}_{pos}|}) = 1$ for $i=1$ and $0$ otherwise to reflect that only the highest ranked document is important to the outcome in push notifications.
We also tested a ``weight capping'' approach for comparison with the loss introduced (see \eqref{eq:weightcapping}) below where $W(\frac{i}{|\mathcal{X}_{pos}|}) = k < 1$ for $i>1$, so that lower ranked positive examples still contribute to the loss, just at a reduced weight.
We treated $k$ as a hyperparameter and include $k=0$ (where it reduces to the original K-OS loss) in the hyperparameter search space.
In \citet{weston2013learning} \eqref{eq:kos} was estimated by sampling due the prohibitively large number of documents. In our case, each candidate set is of moderate size, so we compute the loss exactly.

\subsection{Pseudo-candidate sets} \label{sec:pseudocandidatesets}

Applying pairwise or listwise losses to push notifications is not straightforward due to the previously discussed requirement that only a single document from each candidate set can be sent to the user. Since only one item in the set is labelled, we must find some way to group items for pairwise or listwise comparisons.

The approach we used for all the experiments outlined below was to load a batch of candidates from the training log, and group examples by user type. This approach was used for all pairwise and listwise losses in the experiments. Appendix \ref{appendix:alternativespseudo} outlines some alternative approaches we considered.

\subsection{Expected regret}

Here we introduce the ranking loss that is the primary contribution of this work. % As with many ranking losses, this loss can be viewed as a weighting over pairwise losses.
The motivation of the \textit{expected regret} (ER) loss is to weight pairwise losses by the regret that is expected to be incurred if only this pair was misordered in the ranking. The loss is over pseudo-candidate sets.% With a pairwise hinge the loss is structured as:
\begin{align} \label{eq:er1}
    \ell_{er}(u, \theta) = \sum_{\mathcal{X}_{pos}} \sum_{\mathcal{X}_{neg}} w_{er}(x_{pos}, x_{neg})\times\max(0, 1 - (f_{\theta}(u, x_{pos}) - f_{\theta}(u, x_{neg})))
\end{align}

The weighting of each pairwise loss $w_{er}(x_{pos}, x_{neg})$ is the crucial aspect. We introduce the approach by assuming that for each document $x_{pos}$ we have three additional pieces of information: the probability of the user opening the document $\tilde{y} = p(y=1| x_{pos}, u)$ (note $\tilde{y} \in [0, 1]$ is not binary), the cumulative distribution function for $\tilde{y}$ for this candidate set (we denote as $F(\tilde{y})$) and the number of candidates in the candidate set $n$. Knowledge of $\tilde{y}$ is unrealistic, and we will remove later. However, $F(\tilde{y})$ can be estimated from logged data and $n$ is known.

We compute $w_{er}$ for a pair by using the information to compute the expected regret in CTR (Click Through Rate) incurred if the model misorders only this pair in a candidate set. Since we send only one candidate, the regret incurred for misordering anything but the top ranked candidate is $0$ so this can decomposed into two parts: the probability $x_{pos}$ is the top ranking candidate and the expected regret is $x_{neg}$ is sent rather than $x_{pos}$.

The probability that a candidate with a CTR of $\tilde{y}_{pos}$ is the top ranked candidate in a candidate set of size $n$ is:
\begin{align}
  p_{top}(\tilde{y}_{pos}) = (1 - F(\tilde{y}_{pos}))^{n - 1}
\end{align}
If the document $x_{pos}$ should have been the top-ranked document and we misorder and send the user $x_{neg}$ instead, then we will incur a regret in expected CTR of $\tilde{y}_{pos} - \tilde{y}_{neg}$. Therefore the expected regret of misranking a pair is
\begin{align} \label{eq:expected_regret}
  w_{er}' &= p_{top}(\tilde{y}_{pos})\times (\tilde{y}_{pos} - \tilde{y}_{neg})
  \\
  w_{er} &= \max(w_{er}', k) \label{eq:weightcapping}
\end{align}
and we bound the weights so they cannot be negative or completely ignore any examples.

\subsubsection{Removing the CTR assumption}
\label{sec:remotectr}

The ER loss proposed above requires using $\tilde{y}$ when weighting the pairs. Obviously, assuming knowledge of the latent value of a document is unworkable, if we knew that we would know the perfect ranking. Here we remove that assumption by observing even if the pointwise losses don't provide good ranking performance, for the purposes of weighting pairs, we only need to estimate $\tilde{y}$ approximately. The estimated CTR is used only for weighting, the labels are used for generating pairs so errors in the CTR estimate will result in weighting a pair incorrectly but won't result in the loss ranking a negative document above a positive one.

\label{sec:erpoint}
To avoid doubling the training time, by requiring first training a pointwise loss and then a ER loss, we used the same model and train with both losses simultaneously. Combining pointwise and pairwise losses has previously been successful with logistic regression \citep{sculley2010combined, li2015click}.
However, it was crucial to ensure the losses are approximately compatible between the pairwise and pointwise loss. We used an $\ell_2$ pointwise loss and used the label values $1, -1$ for the $\ell_2$ loss so that for any pair $x_{pos}$, $x_{neg}$ there is a solution that minimizes both the pairwise and pointwise loss. We add these losses together with a hyperparameter $\alpha$ to control their relative contribution to the final loss.

Algorithm \ref{algo:er-no-cheat} puts everything together to show how theses losses can be combined to simultaneously learn online both a CTR estimate, and using this estimate to weight the pairwise loss (on the same model).

\begin{algorithm}
  \begin{algorithmic}
    \State Initialize $\theta$
    \For{training iterations}
      \State Sample pair of examples $(u_{pos}, x_{pos}), (u_{neg}, x_{neg})$
      \State Estimate CTRs $\tilde{y}'_{pos} = f_{\theta}(u_{pos}, x_{pos}),\: \tilde{y}'_{neg} = f_{\theta}(u_{neg}, x_{neg})$.
      \State Compute $w_{er}$ using \eqref{eq:expected_regret} and the estimated values $\tilde{y}'_{pos}, \tilde{y}'_{neg}$.
      \State Compute ER loss $\ell_{er}$ using \eqref{eq:er1}
      \State Sample pointwise sample $u_i, x_i, y_i$.
      \State Compute pointwise loss $\ell_{2}$.
      \State Update $\theta$ to minimize loss 
      \begin{equation}\label{eq:alphaweight}
          \ell = \ell_{er} + \alpha \ell_{2}
      \end{equation}% ($\alpha=0.3$ in our experiments)
    \EndFor
  \end{algorithmic}
  \caption{Pseudo-code of the (unbatched) ER loss.}
  \label{algo:er-no-cheat}
\end{algorithm}

\section{Experiments}
We compared the Expected Regret loss with the pointwise loss, pairwise loss, K-OS-AUC loss. We treated the capping weight $k$ as a hyperparameter for K-OS loss. 
For all losses the same neural network architecture, dataset, and feature pre-processing were used, and models were trained using stochastic gradient descent with Adam optimizer \citep{kingma2014adam} in simulation and plain SGD in the production experiment.

\subsection{Simulation}

We tested Expected Regret loss (along with baselines) in an production experiment (next section) and a simulated experiment using synthetic data which we describe here.

Simulations of recommender systems have gained popularity as systems to facilitate experimental setups beyond simple pointwise losses \citep{rohde2018recogym, ie2019recsim, ie2019reinforcement, mladenov2021recsim}. We used the recsim framework \citep{ie2019recsim} to construct the simulation, and we fit key parameters of the simulation to production data in order to make the simulation more realistic.

We used an extremely simple model of users as being fully described by their user type, sampled from a Categorical distribution matching the production system. Conditioned on the user type a candidate set consisting of 60 documents is sampled at each interaction. Each document has an underlying probability of being opened $\tilde{y} = p(y=1|x, u)$ which is drawn independently and identically from a Beta distribution fit (using maximum likelihood estimation) to the candidate distribution of the production system. For fitting the distributions, the production ranking system estimates were taken as ground truth. Each document has a corresponding set of features $x$, which in our case is a 5 degree projection of $p(y=1|x,u)$ with independent Gaussian noise added to each dimension. Although a very simple model of features, this captures the essence that features provide a model with a noisy estimate of the underlying CTR of the document. A model is used to score the documents and select which one is sent to the user, and a label is generated by sampling from the Bernoulli distribution defined by the latent CTR $\tilde{y}$ associated with the document. For evaluation we used the latent probability of open to compute the regret without any noise due to the label sampling (see section \ref{sec:pushutility}).

\subsection{Data bias}

The ranking models are trained on feedback from users. A well-known problem in ranking is the data bias  \citep{wang2016learning,levine2020offline} issue this introduces, since feedback will only be received on documents that are sent to a user, typically documents which rank highly under an existing ranker. We don't propose to solve this common problem here. Ideally, we might explore by sending the users documents chosen uniformly at random over all possible candidates. However, this would result in a poor user experience, so a common trade-off is $\epsilon$-greedy exploration, where for a small $\epsilon$ fraction of requests documents are sent at random, and for the remainder the top ranked document is sent.

In simulation we test the ranking losses with both ``unbiased'' data, which simulates sending documents to a user chosen at random and a biased production dataset. We simulated the production data bias by ranking documents according to a pointwise ranker (trained on an second set of simulated data) with $\epsilon$-greedy exploration  with $\epsilon$ of $0.14$, approximately the same as the production data. 

\subsection{Online production experiments}

For the production experiment all models were trained offline on the same logged production data which was obtained using $\epsilon$-greedy exploration $\epsilon \approx 0.14$\footnote{This is not the true value of $\epsilon$, rather non-exploration notifications are downsampled so this is a description of the data.} and the production ranking model. % All models were trained on the same data and used the same architecture (other than the loss). Stochastic Gradient Descent was used for optimization and learning rates were tuned for each loss.
The models were tested by serving a subset of Twitter users for push notifications. The experiment ran for approximately two weeks. Users were kept in the same ``bucket'' throughout the experiment, receiving notifications from only one of the models. The hyperparameters of the model can be found in Appendix~\ref{Hyperparam}. 

\section{Results}

The results for the simulated data are in Table \ref{fig:sim-results}. The key finding is that on unbiased data expected regret outperforms all the alternative approaches. However, on biased data expected regret performs slightly worse than a simple pointwise or pairwise loss (but still outperforms the baseline listwise ranking loss K-OS). The poorer performance on biased data is probably due to the the expected regret method weighting high-ranking examples heavily. The data bias already results in over-representing high-ranking examples and this may be resulting in an extreme overweighting on high ranked examples.

\begin{table}
\begin{minipage}[b]{.5\textwidth}
  \centering
\begin{tabular}{lll}
\hline
Model         & unbiased           & gain    \\
\hline
Pointwise     & 0.07624 $\pm$ 0.00006 & 0.0\%   \\
Pairwise      & 0.07623 $\pm$ 0.00006 & 0.01\%  \\
K-OS-AUC      & 0.07679 $\pm$ 0.00016 & -0.72\% \\
Expect Regret & \textbf{0.07602} $\pm$ \textbf{0.00003} & 0.28\% \\
\hline
\end{tabular}
\end{minipage}\qquad
\begin{minipage}[b]{.5\textwidth}
  \centering
\begin{tabular}{lll}
\hline
Model         & biased       & gain   \\
\hline
Pointwise     & 0.07417 $\pm$ 0.00005 & 0.0\%\\
Pairwise      & \textbf{0.07412} $\pm$ \textbf{0.00004} & 0.07\% \\
K-OS-AUC      & 0.08139 $\pm$ 0.00119 & -9.73\% \\
Expect Regret &  0.07429 $\pm$ 0.00008 & -0.16\% \\
\hline
\end{tabular}    
\end{minipage}
  \caption{\textbf{The regret of different ranking losses on simulated data}. The units of regret are arbitrary, and lower is better. Errorbars show SEM from 10 runs. Gains show the \% improvement in the regret compared to the pointwise loss. The left panel is evaluated on unbiased data and the right panel is on biased data.}
  \label{fig:sim-results}
\end{table}

We also compared the methods in a large-scale AB testing experiment on Twitter. Due to the work incurred in production experiments and its poor performance in simulation, we did not include the K-OS loss and its variant. We did compare against pointwise and pairwise ranking models.

Table \ref{table:DDG} shows the results of the production experiment. Despite the issues of bias in simulation, expected regret clearly outperformed other methods in the real-world simulation. It resulted in both a statistically significant improvement in the open rate and users interacted more with the opened documents, indicating the documents being recommended were of more interest to them.

One reason the data bias issue may affect the production experiment less is that the simulated data has symmetric Gaussian noise, which means the performance of a model is very sensitive to precise weighting over the data, whereas this is not true of the production experiments.

\begin{table}[h!]
\centering
\begin{tabular}{|l|c|c|c|}
                              Model                                                      & Open       & Interact \\
\hline
Pointwise                                                                     & 0.0\%           & 0.0\%                    \\
Pairwise                                                                       & 0.04\%          & 0.46\%                 \\
\begin{tabular}[c]{@{}l@{}}Expected regret \end{tabular} & \textbf{0.33\%} & \textbf{2.46\%}           
\end{tabular}
\caption{\textbf{Online experiment results} (\textbf{bold number} for differences $p < 0.01$). The table shows \% gain in open and interaction rate compared with the baseline pointwise model (the pointwise model performance is 0\% by definition). The expected regret model resulted in a higher open rate compared to both a pointwise and pairwise model. Although not directly optimized for, the interact rate, which indicates a user subsequently interacted with a document after opening was also significantly improved. This indicates the improved ranking also resulted in users being more engaged with the content.}
\label{table:DDG}
\end{table}

\section{Conclusion}

In this work we have outlined the characteristics of ranking for push notifications that make it challenging and distinct from other ranking problems. We introduced a novel \textit{expected regret} (ER) ranking loss designed for push notifications, which weights a pairwise loss to minimize the expected regret in the ranking. We compared ER against prior approaches in both a simulation and a real-world production setting. In both cases, we found significant improvements in performance over prior methods.

\bibliographystyle{ACM-Reference-Format}
\bibliography{expected_regret}

\appendix

\section{Alternative approaches to pseudo-candidate sets}
\label{appendix:alternativespseudo}

As outlined in section \ref{sec:pseudocandidatesets}, in order to apply pairwise or listwise losses we grouped candidates set to the user into pseudo-candidate sets. Here we discuss two alternative approaches to constructing pseudo-candidate sets we considered.

The fundamental challenge is that, due to requirement to send a single document from a candidate set, the log data consists of user candidate pairs $(u_1, x_1, y_1), (u_2, x_2, y_2), ...$ which differ in both the user attributes $u$ and document $x$ and user responses $y$. However, when ranking, we are considering various documents for a single user, so the candidate set will have the same user attributes for every document $(u, x_1), (u, x_2), ...$. We can't get this log because we can only send the user a single notification at a time, so we do not have the labels $y$ for the other documents.

One approach would be to load a batch of examples and treat them all as a candidate set. However, one of the most important features that predict if a user will open a notification is their ``user type,'' which classifies a user by how actively they engage with the platform. Grouping all users in a batch would encourage the model to learn to simply pick out very active users and rank documents sent to them highly which would not work well for online ranking, where candidate documents are all being sent to the same user.

Another approach we tried is to group all notifications sent to the same user over some period of time together. However, we found this performed very poorly (data not shown) because the model learns to ``cheat'' by making use of user features that change based on their responses (e.g.\ features indicating if the user has been active on the platform recently). Concretely, even though document $x_1$ and $x_2$ were sent the same user, some user attributes can change in the interval between notifications so the log data looks like $(u, x_1), (u', x_2)$ where $u'$ is the same user with some features changed. The model can exploit those feature changes to predict which document was opened in a pairwise loss by just focusing on the user features. This results in good offline performance on the pseudo-candidate sets, but abysmal performance when used for ranking over real candidate sets.

\section{Hyperparameters}\label{Hyperparam}

For both simulation and online experiment, we use $k=0.001$ for the weight capping in \eqref{eq:weightcapping} and $\alpha=0.3$ for the weight of pointwise loss \eqref{eq:alphaweight}.

Simulation: For the simulation, we use a single set of neural network structure and hyperparameters for a fair comparison. The simple MLP has a size of $(64,32)$ and sigmoid activation functions. We use the Adam optimizer \citep{kingma2014adam} with learning rate $0.001$ and rest parameters are default by the Keras API \citep{chollet2015keras}. The batch size for all methods is $512$ and the early stopping patience is $5$. 

Online Experiment:
 On top of the extracted features, we used a three-layer MLP with size $(256, 128, 64)$ and the activation function is LeakyReLU \citep{xu2015empirical}. We used a SGD optimizer \citep{ruder2016overview} with momentum $0.99999$. 
 For each method we use cross-validation to choose an optimal learning rate (from $\{1e^{-1}, 1e^{-2},1e^{-3},1e^{-4} \}$) and batch size (from $\{64, 128, 256, 512, 1024 \}$).

\end{document}